\documentclass[a4paper,11pt]{article}
\usepackage{pos}

\usepackage{bibspacing}
\title{Probing VHE gamma-ray emission from GW events with H.E.S.S.}
 \ShortTitle{GWs with H.E.S.S.}

\author*[a]{Halim Ashkar}
\author[b]{Mathieu de Bony de Lavergne}
\author[b]{Francois Brun}
\author[a]{Stephen Fegan}
\author[c]{Ruslan Konno}
\author[c]{Stefan Ohm}
\author[c]{Heike Prokoph}
\author[b]{Fabian Sch\"ussler}
\author[c]{Sylvia J Zhu}

\affiliation[a]{Laboratoire Leprince-Ringuet, École Polytechnique, CNRS, Institut Polytechnique de Paris, F-91128 Palaiseau, France}

\affiliation[b]{IRFU, CEA, Université Paris-Saclay, Gif-sur-Yvette, France}

\affiliation[b]{DESY, D-15738 Zeuthen, Germany}

\onbehalf{on behalf of the H.E.S.S. collaboration}

\emailAdd{contact.hess@hess-experiment.eu}

\abstract{Gravitational wave (GW) events, particularly those connected to the merger of compact objects such as neutron stars, are believed to be the primary source of short gamma-ray bursts. To explore the very high energy (VHE) component of the emission from these events, the H.E.S.S. collaboration has dedicated a substantial effort and observing time to follow up on these events. During the second and third GW observing runs, H.E.S.S. was the first ground-based instrument to observe the GW170817 binary neutron star merger. In addition, H.E.S.S. followed four binary black hole mergers. The data acquired by H.E.S.S. was used to constrain the VHE emission from these events for the first time. H.E.S.S. also monitored the GW170817 source for approximately 50 hours and obtained limits that constrained the magnetic field in the merger remnant to $> 24  \mu G$. As the fourth GW observing run (O4) approaches, the H.E.S.S. collaboration has allocated significant observation time to the follow-up of GW events. This contribution provides an overview of the science results derived from the H.E.S.S. follow-up of GW events, a technical overview of the GW follow-up strategies for O4, and an update on H.E.S.S. activities during O4.}

\ConferenceLogo{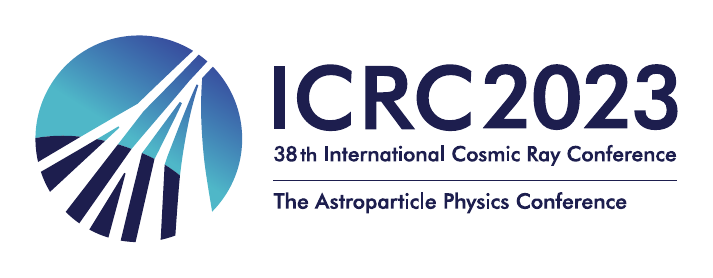}

\FullConference{%
38th International Cosmic Ray Conference (ICRC2023)\\
  26 July - 3 August, 2023\\
  Nagoya, Japan}

\begin{document}
\maketitle

\section{Introduction}
\label{sec:introduction}
Gravitational Wave (GW) events, such as the merger of compact objects, are a source of interest to high-energy astronomers for probing the highest energy photons in the gamma-ray ray domain. Mergers including neutron stars are known to be responsible for a significant portion of gamma-ray bursts (GRBs), notably short GRBs. Probing the very high energy (VHE) emission from these cataclysmic events, with imaging atmospheric telescopes (IACTs) such as the High Energy Stereoscopic System (H.E.S.S.)  brings information on the: non-thermal emission processes creating the highest energy photons, particle acceleration mechanisms in extreme magnetic fields and properties of the merger remnant. 

GW event detection suffers from poor localization as their localization regions can span tens to thousands of degrees in the sky. Due to their low duty cycle (around 10\%), IACTs struggle with the low latency follow-up of such targets of opportunity (ToO) as they can only observe in quasi-total darkness. However, due to their medium-to-large field of view (FoV), they have an advantage over small FoV instruments as they can probe large regions in the sky at once. Moreover, they have a superior sensitivity to large-FoV space-based gamma-ray surveying instruments. 

In this contribution, we present an overview of the H.E.S.S. GW program since its beginning before the second GW observing run O2 until today. In Sec.~\ref{sec:strategy} the H.E.S.S. GW follow-up observation strategies are described. In Sec.~\ref{sec:O23Observations} the H.E.S.S. follow-up observations of GW events during O2 and O3 and their implications is presented. Finally, Sec.~\ref{sec:O4Observations} outlines the preparations and the H.E.S.S. activities during the fourth observing run O4.

\section{H.E.S.S. GW follow-up strategy}
\label{sec:strategy}
To efficiently cover GW events and increase the chances of catching the source (and the VHE counterparts) as fast as possible in the large GW localization regions, the H.E.S.S. collaboration has dedicated extensive efforts to the development and optimization of GW follow-up strategies. 

\subsection{Science cases}
The aim of the H.E.S.S. follow-up observations of GW events is to probe VHE emission issued from particle acceleration in GRBs from the merger of compact objects. Binary neutron star (BNS) mergers are prime candidates for producing such GRBs. This has been proven in 2017 by the dual detection of GW170817 and GRB 170817~\citep{GW170817MM}. Not only BNS mergers are expected to produce GRBs but also black hole-neutron star (BHNS) merger~\citep{Foucart_2020_NSBHreview}. Therefore, BNS, BHNS mergers, and MassGAP events (an object falling in the maximum neutron star and the minimum black hole mass gap) are considered in one science case and are given the same priority. Since these cases have a high scientific yield, the follow-up criteria are loose for this science case. A follow-up observation that fulfills the H.E.S.S. observation and visibility conditions (darkness or moderate moonlight with a maximum 60 deg zenith angle) with coverage exceeding 10\% of the localization region within 24 hours of the event, is considered interesting enough for H.E.S.S. to spend observing time on it. 

In the case of BBH mergers, electromagnetic emission is not highly anticipated. However, a hint of a high energy transient coincident with GW150914 has previously sparked some interest in the community~\citep{2016ApJ...826L...6C}. Moreover, some extreme scenarios of BBH mergers predict the emission of electromagnetic waves~\citep{Perna2019_BBHEM, Murase2016_BBHEM, Kotera2016_BBHEM, Martin2018_BBHEM}. Since it is important to verify these hypotheses, the H.E.S.S. collaboration also considers BBH mergers and follows GW events emanating from such sources. In that case, good coverage of the localization region is required, to maximize the chances of covering the GW source. This allows us to efficiently reject the emission of VHE gamma rays and to place stringent upper limits in a non-detection case. Therefore, in addition to having good observation and visibility conditions, the coverage requirement to follow BBH mergers is above 50\%. 

Burst alerts are GW events from non-modeled sources such as nearby asymmetric supernovae. The sources of burst alerts can be extremely interesting for IACTs, due to their close distances. However, the false alert rate of Burst events is higher than in compact binary mergers. They fall in the domain of exploratory searches and their priority lies between neutron stars mergers and BBH mergers. The requirement for Burst follow-up with H.E.S.S. is a coverage of more than 20\%.  

All alerts should have a probability of Terrestrial origin lower than 50\% in order to trigger a H.E.S.S. response.

\subsection{Observation strategy}
To efficiently cover GW event localization regions, the H.E.S.S. collaboration developed 2 strategies to optimize the scheduling of GW follow-up observations~\citep{hessGWFollowuo_technical}. The first strategy (2D strategy) only takes into consideration the probability information contained in the first layer of the probability maps provided by LVKC. Taking advantage of the H.E.S.S. FoV, the probability is integrated. The 90\% GW localization region is divided into smaller regions representing the telescope FoV. At a given observation time fulfilling H.E.S.S. observation conditions, the FoV covering the region with the largest integrated probability of hosting the event has the highest priority to be observed at this given time. The regions are then masked. In the next observation window, the same integrated probability computation is repeated with the remaining regions. 

The second strategy (3D) uses in addition the distance information contained in the other 3 layers of the probability maps and the distribution of galaxies in the local Universe using the GLADE galaxy catalog~\citep{dalya2018glade1}. Here, the galaxies inside the GW localization region, at the distances of the event are assigned a probability. As for the 2D case, several regions in the sky are tested, with the difference that this time the probability of the galaxies hosting the event is integrated inside the H.E.S.S. FoV. The galaxies in the highest probability FoV region are observed. For the next observation window, the same procedure is repeated with the remaining regions/galaxies. 

The choice of a 2D or a 3D strategy depends on the GW event localization information. Galaxy catalogs are incomplete at large distances. Taking that into consideration, a threshold of 150 Mpc was applied for the choice of strategy during O3. All events with an average distance below 150 Mpc are observed with a 3D strategy. All events beyond this limit are observed with a 2D strategy. Moreover, since galaxy catalogs are also incomplete around the galactic plane, all GW events that have localization maps with the hotspot located around the galactic plane are observed with a 2D strategy. 

The H.E.S.S. response and processing of GW alerts is automatized. Human intervention is only needed for checks and irregular updates. Experts on call are provided with tools based on the Tilepy\footnote{\url{https://github.com/astro-transients/tilepy}}~\citep{tilepy_icrc2023} library in case such intervention is needed. A possible case requiring human intervention is the modification of the GW follow-up schedule taking into consideration observations occurring previous to an incoming update to avoid overlap.

All GW events fulfilling the requirements described above and observable within 24 hours are followed. If the GW event occurs outside observation hours, afterglow-mode observations are scheduled for the upcoming night. In that case, observations will be updated following incoming LVKC updates and can be human-vetted. In the case of a GW event occurring during observation hours and given its priority, a prompt-observations mode is triggered and the telescopes will slew automatically towards the highest probability (using a 2D or 3dD strategy) region visible at the time of the arrival of the alert. The condition to trigger such a response is that the probability of hosting the event covered by the prompt observation be higher than 5\% in this single observation. In addition to the rest of the observations, the overall coverage depends on the science case as mentioned above. EarlyWaning, Preliminary, and Initial GW alerts\footnote{\url{https://emfollow.docs.ligo.org/userguide/content.html}} are subject to this prompt-mode response. Updates are only subject to afterglow-mode response. 

During the observation, if the real-time analysis finds a hotpost indicating a significant excess of VHE gamma-rays coming from a region that is not associated with any known source, the hotspot is observed again as it might be a GW VHE counterpart candidate. 

\section{H.E.S.S. GW follow-up during O2 and O3}
\label{sec:O23Observations}
The first H.E.S.S. follow-up of GW occurred during O2 on GW170502. During O3, H.E.S.S. observed GW200105. However, in these cases the H.E.S.S. coverage of the localization region was low as in the first case the localization region was large and it was the first H.E.S.S. trial and in the second case, the localization region shifted significantly after an event update. Therefore, the follow-up of these events will not be featured here. Five more events were observed with H.E.S.S. with a good coverage exceeding 50\%. These events are the BNS merger GW170817 and the BBH mergers GW170814, GW190512, GW190728, and GW200224.

\subsection{GW170817 observations with H.E.S.S.}
In August 2017, GW170817 was detected emanating from a BNS merger followed by a short GRB detected 2 seconds later by Fermi-GBM and INTEGRAL. H.E.S.S. used the updated localization region, distributed a few minutes before the beginning of astronomical dark time on the H.E.S.S. site in Namibia, and derived an optimized follow-up schedule that contained 3 positions to be observed during the 1.5 hours-long visibility window. Six hours after the observations started, the optical counterpart~\citep{Hjorth2017_170817redshift} was discovered in the NGC 4993 galaxy. It turned out that the first H.E.S.S. observation on GW170817 covered NGC 4993, making the data acquired by H.E.S.S. the earliest ground-based data taken on the source. The following night, H.E.S.S. observed the source for a total of 3.2 hours and continued monitoring for several days afterward. The data analysis did not show any significant detection of VHE gamma-rays in the direction of NGC 4993~\citep{HESS170817}. These H.E.S.S. observations permitted to place the first stringent constraint on VHE emission from BNS mergers. The merger occurred at an off-axis viewing angle. The H.E.S.S. limits, later on, helped constrain the off-axis viewing angle of the jet to a value larger than 15 deg~\citep{Murase_2018}. 
\begin{figure}
  \centering
    \includegraphics[width=0.7\textwidth]{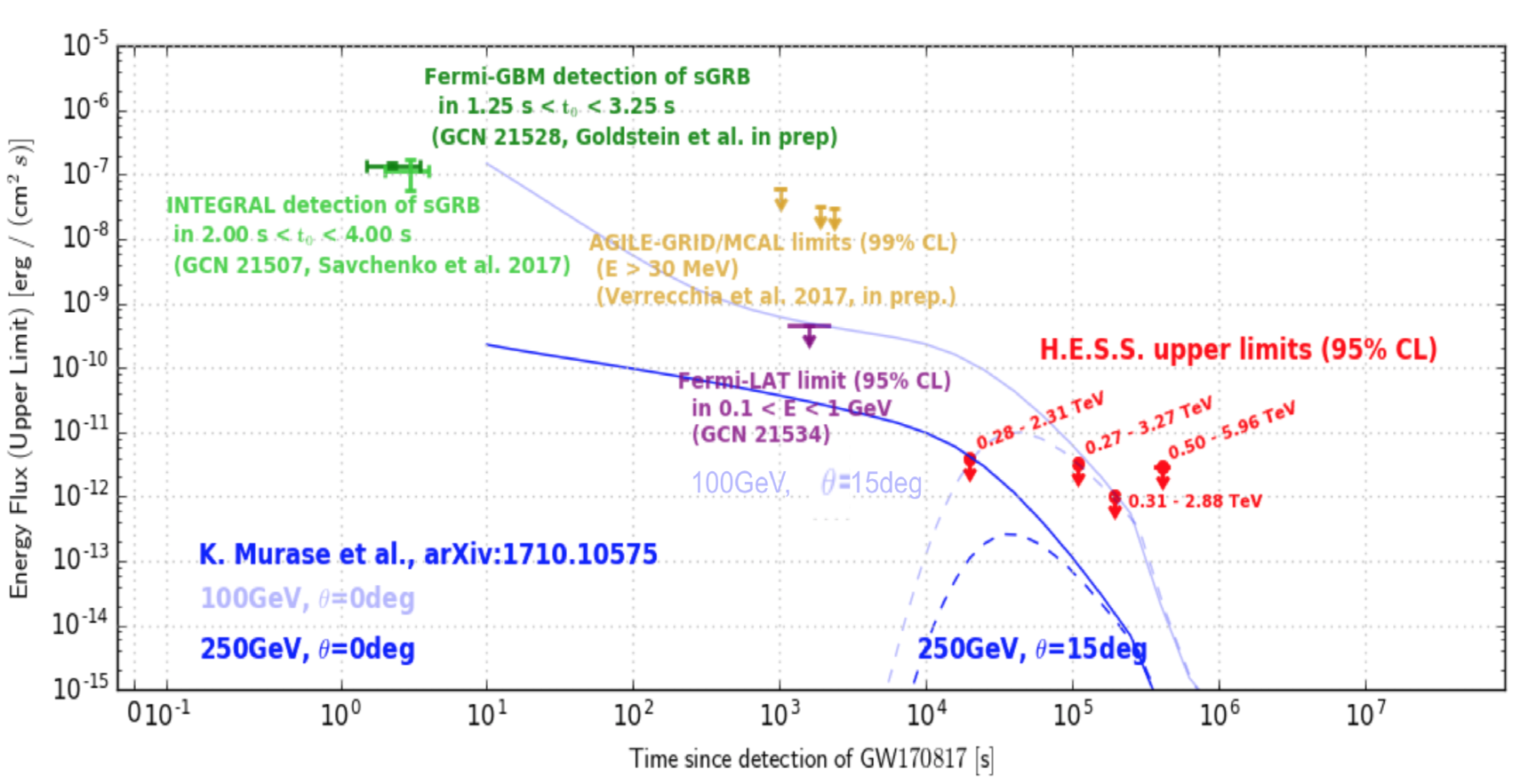}
    \caption{Gamma-ray lightcurve generated by external inverse Compton radiation for emission at 100 and 250 GeV considering viewing angles of 0 and 15 degrees with upper limits on the VHE spectrum derived from H.E.S.S. observation on GW170817. From~\cite{seglararroyo:tel-02936503}, adapted from~\cite{Murase_2018}.}
    \label{fig:hess_lightcurve_model}
\end{figure}

Nine and sixteen days after the merger, the radio and X-ray synchrotron emission from the source started rising as the opening angle of the jet increased. The acceleration of particles in the merger remnants is believed to be suitable for synchrotron self-Compton (SSC) emission, where the high-energy electrons accelerated in the magnetic field upscatter the synchrotron photons created by the same electron population to VHE energies. H.E.S.S. performed a long-term follow-up observation campaign on SSS17a~\citep{HESS170817_deep}. The observation campaign gathered 53.9 hours of data over $\sim$5 months. The analysis did not show any significant VHE emission. The upper limits on the SSC emission are transformed into limits on the strength of the magnetic field of the merger remnant. The synchrotron component brings information on the energy density of the electrons and the magnetic field, but cannot disentangle the two. The SSC component can break the ambiguity. The magnetic field is constrained to $ B > 24~ \mu G$ for an off-axis relativistic jet.
\begin{figure}
  \centering
    \includegraphics[width=0.50\textwidth]{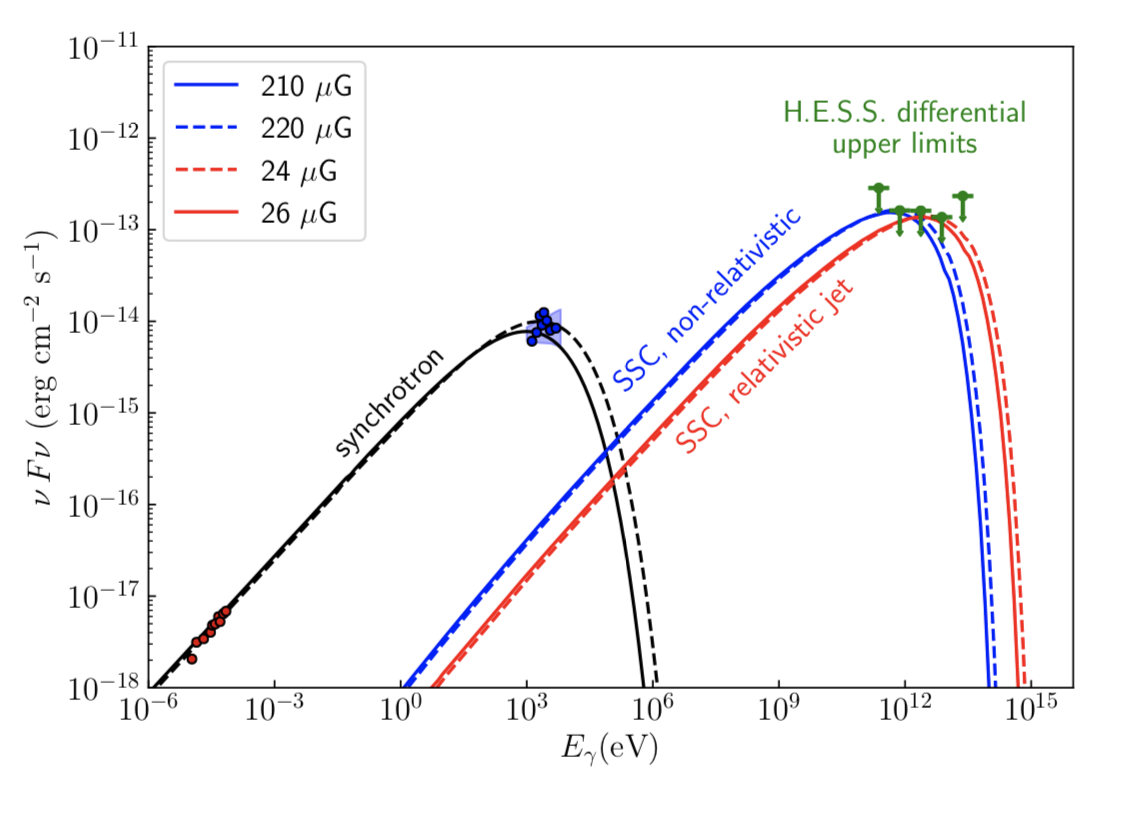}
    \caption{Spectral energy distribution of EM170817 for the non-relativistic (blue lines) and relativistic (red lines) scenarios. The blue and red dots correspond respectively to the X-ray and radio measurements. The green dots represent the H.E.S.S. derived upper limits. The solid and dashed lines correspond respectively to the minimum and maximum X-ray emission. From~\citep{HESS170817_deep}.}
    \label{fig:EM170817_S_IC_PRES_1}
\end{figure}

\subsection{BBH merger observations with H.E.S.S.}
\begin{figure*}[!ht]
  \centering
    \includegraphics[width=0.5\textwidth]{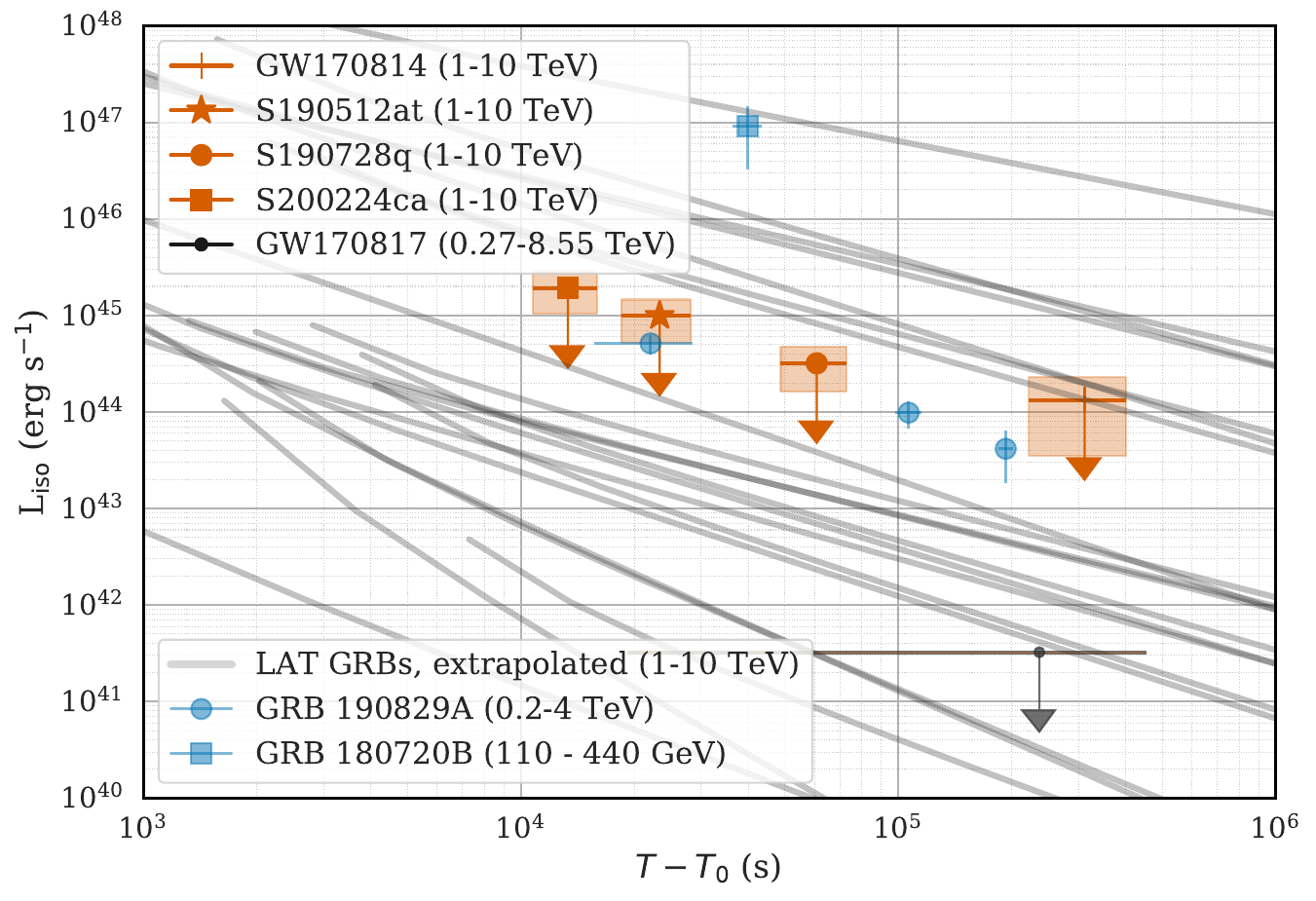}
    \caption{Mean (orange points) and standard deviation (orange bands) of the per-pixel luminosity upper limit maps for the BBH events. They are compared to luminosity extrapolation of Fermi-LAT GRBs (grey lines) with known redshift, The luminosity from H.E.S.S. detected VHE GRBs and to the H.E.S.S. upper limit on GW170817 (black)~\cite{HESS170817}. All five GW upper limits are calculated assuming an intrinsic $E^{-2}$ spectrum, although the upper limit for GW170817 is calculated with a slightly different energy range. From~\cite{Abdalla_2021}.}
\label{fig:luminosity_comparisons}
\end{figure*}

In addition to GW170817, H.E.S.S. also followed up on four BBH mergers: GW170814, GW190512, GW190728, and GW200224. With the 2D strategy, the H.E.S.S. coverage exceeded 50\% for all these events~\citep{Abdalla_2021}. Since no counterparts were detected for these events, the VHE analysis concentrated on searching for VHE signals in all the areas observed. No significant VHE emission is found. Upper limit maps on the VHE energy flux from these events are published. In addition, taking into consideration the GW event distance estimation, the VHE luminosity from these events is constrained\footnote{\url{https://www.mpi-hd.mpg.de/hfm/HESS/pages/publications/auxiliary/2021_BBH_O2_O3/}}. These upper limits are then compared to extrapolation of Fermi-LAT detected GRBs in the VHE domain as shown in Fig.~\ref{fig:luminosity_comparisons}. These comparisons show that the H.E.S.S. observations constrain well the VHE luminosity of these events since the limits placed coincide with the bulk of the GRB extrapolated VHE luminosity. To significantly increase the chances of detecting a VHE counterpart, the main focus should be on getting earlier observations from the time of the merger, something that the H.E.S.S. collaboration has no control over beyond optimizing observation strategies. However, this can be achieved with an increased rate of detection and better localization in upcoming GW observing runs.

\section{H.E.S.S. GW follow-up during O4}
\label{sec:O4Observations}
\begin{figure}[!ht]
  \centering
  \begin{minipage}[b]{0.5\textwidth}
    \includegraphics[width=\textwidth]{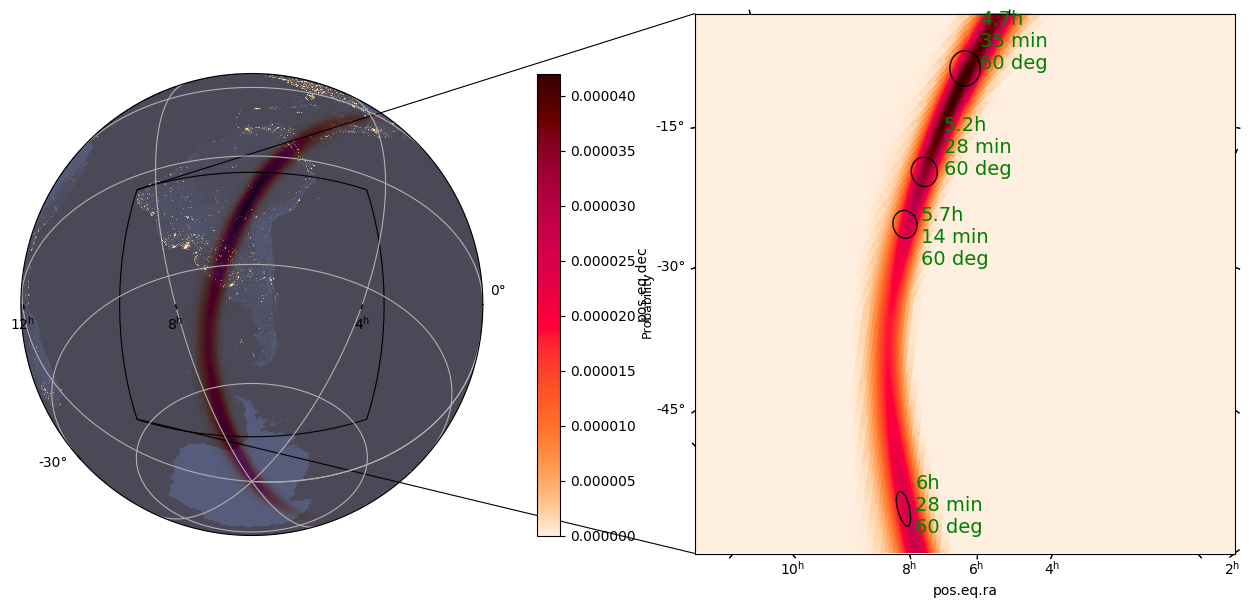}
  \end{minipage}
  \hfill
  \begin{minipage}[b]{0.5\textwidth}
    \includegraphics[width=\textwidth]{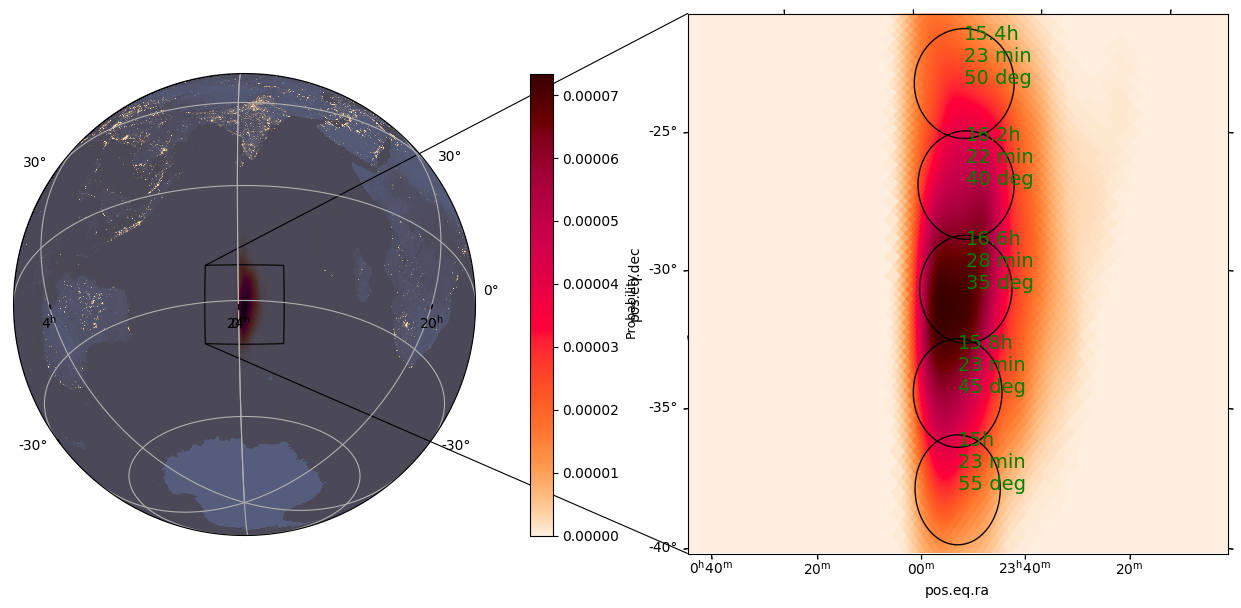}
    \end{minipage}
      \hfill
     \begin{minipage}[b]{0.5\textwidth}
    \includegraphics[width=\textwidth]{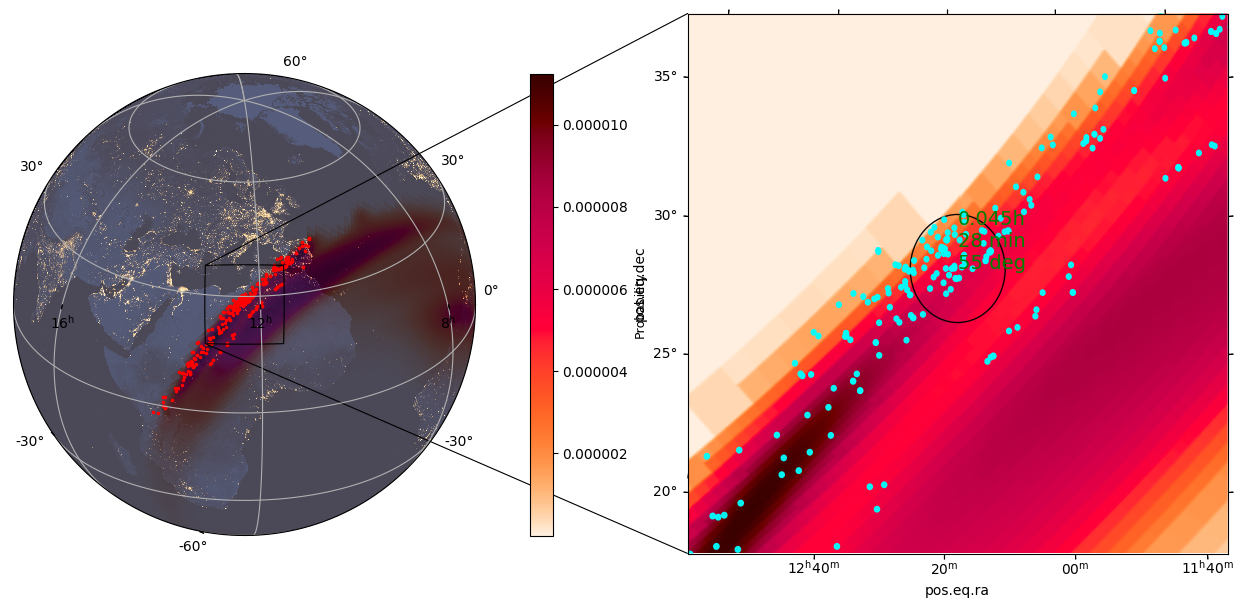}
  \end{minipage}
  \hfill
  \caption{H.E.S.S observations of GW candidate events  (from top to bottom) S230518H, S230528ay and S230615az. On the right the GW localization region is displayed with the Earth in the background at the time of the first H.E.S.S. observation. On the right, the H.E.S.S observations are indicated in a dark circle with the delay, the duration, and the zenith angle. The dots on the bottom plot represent the galaxies at the distance of the event.}
  \label{fig:GW_GESS_O4}
\end{figure}

The lessons learned in the first three GW observing run on the importance of the VHE domain in the search for GW electromagnetic counterparts lead to putting GW observations as the top priority of the H.E.S.S. collaboration. For O4, the collaboration allocated as much as 20\% of its observing time for GW follow-ups. The strategies used in O2 and O3 have been adapted for O4. These adaptations mostly comprised updating the code to the new changes adopted by the LVKC collaboration and the different brokers to the GW VoEvents. The most notable change is the increase of the horizon at which a 3D strategy is used. The value was increased from 150 Mpc to 300 Mpc due to significant improvements in the GLADE catalog. To read the catalog faster, most objects far away from this horizon are removed. Finally, to test the H.E.S.S. response to GW alerts before the start of O4, firedrills consisting in injecting fake GW alerts in the system were performed. These alerts differ from the Mock alerts sent by LVKC as these ones would trigger a telescope active response. The Mock alerts are used to continuously monitor the system and the processing of alerts. However, they do not trigger telescope response as they are marked as "Test" alerts by the H.E.S.S. Too response system~\citep{hessTOOsystem}. The alerts used in the firedrills are based on the GW170817 event and successfully triggered the telescope afterglow and prompt-mode responses. With the firedrills the whole chain of processing from the reception of the alert to the real-time analysis was tested and debugged. These firedrills will be complemented through O4 by expert-on-call training sessions. 

During O4 and until July 2023, H.E.S.S. observed three GW events presented in~\ref{fig:GW_GESS_O4} with a coverage of $\sim$ 10\%, $\sim$ 70\%, and $\sim$ 10\%, respectively. S230518h is the first one received during the engineering run period. It is flagged as 86\% BHNS merger and false alarm rate (FAR) of 1 per 98.463 years and is followed using a mix of 2D and 3D strategies. S230528ay is the second one and it is a Burst alert with a FAR of 22.193 per year and is followed using a 2D strategy. S230615az is the third one and is flagged as an 85\% BNS merge and FAR of 4.7 per year and is followed using a 3D strategy. Although these events are of low significance, they help commission the H.E.S.S. GW follow-up program for 04. Notably, S230516az was received during observation hours. H.E.S.S. successfully triggered a prompt automatic response to GW alerts. During O3, S200224 was received during the night but the prompt reaction could not be tested because the telescopes were parked due to rain. S230615az GW event occurred on 2023-06-15 at 17:50:08. The notice was distributed at 17:50:34 and received by the H.E.S.S. Too response system at 17:50:41. By 17:51:30 the best position that could be observed at the time was  computed with a 3D strategy and distributed to the shifters, the expert on call and forwarded to the telescopes confirming that the time of computation is less than a minute as indicated in~\cite{hessGWFollowuo_technical}. We note that the localization region for this event is significantly large making the processing time longer than the average. The region covered by H.E.S.S. represents $\sim$10\% of the total galaxy probability. This is due to the fact that the algorithms targeted a region with a relatively high concentration of galaxies. The telescopes slewed to this position and started observations at 17:52:02 with a total delay of 114 seconds accounting for the distribution of the alert, the H.E.S.S. processing, and the telescopes slewing.

\bibliographystyle{JHEP}
\bibliography{main}

\providecommand{\href}[2]{#2}\begingroup\raggedright\begin{thebibliography}{10}

\bibitem{GW170817MM}
B.P.~{Abbott}, R.~{Abbott}, T.D.~{Abbott} et~al.,
  ~\href{https://doi.org/10.3847/2041-8213/aa91c9}{\emph{ApJL} {\bfseries 848}
  (2017) L12} [\href{https://arxiv.org/abs/1710.05833}{{\ttfamily
  1710.05833}}].

\bibitem{Foucart_2020_NSBHreview}
F.~Foucart, ~\href{https://doi.org/10.3389/fspas.2020.00046}{\emph{Front.
  Astron. Space Sci.} {\bfseries 7} (2020) 46}.

\bibitem{2016ApJ...826L...6C}
V.~{Connaughton}, E.~{Burns}, A.~{Goldstein} et~al.,
  ~\href{https://doi.org/10.3847/2041-8205/826/1/L6}{\emph{ApJL} {\bfseries
  826} (2016) L6} [\href{https://arxiv.org/abs/1602.03920}{{\ttfamily
  1602.03920}}].

\bibitem{Perna2019_BBHEM}
R.~{Perna}, D.~{Lazzati} and W.~{Farr},
  ~\href{https://doi.org/10.3847/1538-4357/ab107b}{\emph{ApJ} {\bfseries 875}
  (2019) 49} [\href{https://arxiv.org/abs/1901.04522}{{\ttfamily 1901.04522}}].

\bibitem{Murase2016_BBHEM}
K.~{Murase}, K.~{Kashiyama}, P.~{M{\'e}sz{\'a}ros} et~al.,
  ~\href{https://doi.org/10.3847/2041-8205/822/1/L9}{\emph{ApJL} {\bfseries
  822} (2016) L9} [\href{https://arxiv.org/abs/1602.06938}{{\ttfamily
  1602.06938}}].

\bibitem{Kotera2016_BBHEM}
K.~{Kotera} and J.~{Silk},
  ~\href{https://doi.org/10.3847/2041-8205/823/2/L29}{\emph{ApJL} {\bfseries
  823} (2016) L29} [\href{https://arxiv.org/abs/1602.06961}{{\ttfamily
  1602.06961}}].

\bibitem{Martin2018_BBHEM}
R.G.~{Martin}, C.~{Nixon}, F.-G.~{Xie} et~al.,
  ~\href{https://doi.org/10.1093/mnras/sty2178}{\emph{MNRAS} {\bfseries 480}
  (2018) 4732} [\href{https://arxiv.org/abs/1808.06023}{{\ttfamily
  1808.06023}}].

\bibitem{hessGWFollowuo_technical}
H.~Ashkar, F.~Brun, M.~Füßling et~al.,
  ~\href{https://doi.org/10.1088/1475-7516/2021/03/045}{\emph{JCAP} {\bfseries
  2021} (2021) 045}.

\bibitem{dalya2018glade1}
G.~D{\'{a} }lya, G.~Galg{\'{o}}czi, L.~Dobos et~al.,
  ~\href{https://doi.org/10.1093/mnras/sty1703}{\emph{Monthly Notices of the
  Royal Astronomical Society} {\bfseries 479} (2018) 2374}.

\bibitem{tilepy_icrc2023}
F.~{Sch\"ussler}, H.~{Ashkar}, M.~{de Bony de Lavergne} et~al., ~ in \emph{38th
  International Cosmic Ray Conference. July 26 - Aug 3, 2023. Nagoya},
  vol.~PoS(ICRC2023)1469, July, 2023.

\bibitem{Hjorth2017_170817redshift}
J.~{Hjorth}, A.J.~{Levan}, N.R.~{Tanvir} et~al.,
  ~\href{https://doi.org/10.3847/2041-8213/aa9110}{\emph{ApJL} {\bfseries 848}
  (2017) L31} [\href{https://arxiv.org/abs/1710.05856}{{\ttfamily
  1710.05856}}].

\bibitem{HESS170817}
H.~{Abdalla}, A.~{Abramowski}, F.~{Aharonian} et~al.,
  ~\href{https://doi.org/10.3847/2041-8213/aa97d2}{\emph{ApJL} {\bfseries 850}
  (2017) L22} [\href{https://arxiv.org/abs/1710.05862}{{\ttfamily
  1710.05862}}].

\bibitem{Murase_2018}
K.~Murase, M.W.~Toomey, K.~Fang et~al.,
  ~\href{https://doi.org/10.3847/1538-4357/aaa48a}{\emph{ApJ} {\bfseries 854}
  (2018) 60}.

\bibitem{seglararroyo:tel-02936503}
M.~Seglar-Arroyo, \emph{{Studying the origin of cosmic-rays : Multi-messenger
  studies with very-high-energy gamma-ray instruments}}, theses,
  {Universit{\'e} Paris Saclay (COmUE)}, Sept., 2019.

\bibitem{HESS170817_deep}
H.~{Abdalla}, R.~{Adam}, F.~{Aharonian} et~al.,
  ~\href{https://doi.org/10.3847/2041-8213/ab8b59}{\emph{ApJL} {\bfseries 894}
  (2020) L16} [\href{https://arxiv.org/abs/2004.10105}{{\ttfamily
  2004.10105}}].

\bibitem{Abdalla_2021}
H.~Abdalla, F.~Aharonian, F.A.~Benkhali et~al.,
  ~\href{https://doi.org/10.3847/1538-4357/ac2e04}{\emph{ApJ} {\bfseries 923}
  (2021) 109}.

\bibitem{hessTOOsystem}
C.~Hoischen, M.~Fü{\ss}ling, S.~Ohm et~al.,
  ~\href{https://doi.org/10.1051/0004-6361/202243092}{\emph{A\&A} {\bfseries
  666} (2022) A119}.

\end{thebibliography}\endgroup


%
%
%

\section*{Full Author List: H.E.S.S. Collaboration}
\scriptsize
\noindent
F.~Aharonian$^{1,2,3}$, 
F.~Ait~Benkhali$^{4}$, 
A.~Alkan$^{5}$, 
J.~Aschersleben$^{6}$, 
H.~Ashkar$^{7}$, 
M.~Backes$^{8,9}$, 
A.~Baktash$^{10}$, 
V.~Barbosa~Martins$^{11}$, 
A.~Barnacka$^{12}$, 
J.~Barnard$^{13}$
% ...

\medskip

\noindent
$^{1}$Dublin Institute for Advanced Studies, 31 Fitzwilliam Place, Dublin 2, Ireland\\
$^{2}$Max-Planck-Institut f\"ur Kernphysik, P.O. Box 103980, D 69029 Heidelberg, Germany\\
$^{3}$Yerevan State University,  1 Alek Manukyan St, Yerevan 0025, Armenia\\
$^{4}$Landessternwarte, Universit\"at Heidelberg, K\"onigstuhl, D 69117 Heidelberg, Germany\\
$^{5}$IRFU, CEA, Universit\'e Paris-Saclay, F-91191 Gif-sur-Yvette, France\\
$^{6}$Kapteyn Astronomical Institute, University of Groningen, Landleven 12, 9747 AD Groningen, The Netherlands\\
$^{7}$Laboratoire Leprince-Ringuet, École Polytechnique, CNRS, Institut Polytechnique de Paris, F-91128 Palaiseau, France\\
$^{8}$University of Namibia, Department of Physics, Private Bag 13301, Windhoek 10005, Namibia\\
$^{9}$Centre for Space Research, North-West University, Potchefstroom 2520, South Africa\\
$^{10}$Universit\"at Hamburg, Institut f\"ur Experimentalphysik, Luruper Chaussee 149, D 22761 Hamburg, Germany\\
$^{11}$Deutsches Elektronen-Synchrotron DESY, Platanenallee 6, 15738 Zeuthen, Germany\\
$^{12}$Obserwatorium Astronomiczne, Uniwersytet Jagiello{\'n}ski, ul. Orla 171, 30-244 Krak{\'o}w, Poland\\
$^{13}$Department of Physics, University of the Free State,  PO Box 339, Bloemfontein 9300, South Africa\\
% ...

\end{document}